\documentclass[aps,pra,onecolumn,groupedaddress,citesort]{revtex4-2}
\usepackage{mathrsfs}
\usepackage{amsfonts}
\usepackage{amstext}
\usepackage{amsmath}
\usepackage{amssymb}
\usepackage{bm}
\usepackage{bm}
\usepackage{bbm}
\usepackage[dvips]{graphicx}
\def\qed{\leavevmode\unskip\penalty9999 \hbox{}\nobreak\hfill
     \quad\hbox{\leavevmode  \hbox to.77778em{%
              \hfil\vrule   \vbox to.675em%
               {\hrule width.6em\vfil\hrule}\vrule\hfil}}
     \par\vskip3pt}

\usepackage{amssymb}
\usepackage{graphicx}
\usepackage{graphics}
\usepackage{amsmath}
\usepackage{amsthm}
\usepackage{color}
\usepackage{dsfont}
\definecolor{darkred}  {rgb}{0.5,0,0}
\definecolor{darkblue} {rgb}{0,0,0.5}
\definecolor{darkgreen}{rgb}{0,0.5,0}
\usepackage{hyperref}
\hypersetup{
	pdftitle = {QRT Proposal},
	pdfauthor = {},
	colorlinks = true,
	urlcolor  = blue,         
	linkcolor = red,     
	citecolor = blue,    
	filecolor = darkred       
}
\usepackage{mathtools}
\def\ra{\rangle}
\def\la{\langle}
\def\bb{\mathbb}

\newtheorem{pro}{Proposition}

\newcommand{\bea}{\begin{eqnarray}}
\newcommand{\eea}{\end{eqnarray}}
\newcommand{\be}{\begin{equation}}
\newcommand{\ee}{\end{equation}}
\newcommand{\ba}{\begin{equation}\begin{aligned}}
\newcommand{\ea}{\end{aligned}\end{equation}}

\newcommand{\beax}{\begin{eqnarray*}}
\newcommand{\eeax}{\end{eqnarray*}}
\newcommand{\bex}{\begin{equation*}}
\newcommand{\eex}{\end{equation*}}

\newtheorem{definition}{Definition}
\theoremstyle{remark}
\newtheorem{remark}{Remark}

\newcommand{\mE}{\mathcal{E}}
\newcommand{\mF}{\mathcal{F}}

\newcommand{\mH}{\mathcal{H}}

\newcommand{\mS}{\mathcal{S}}

\newcommand{\tr}{{\rm Tr}}

\def\>{\rangle}
\def\<{\langle}

\begin{document}


\preprint{APS/123-QED}
\title{When is a Genuine Multipartite Entanglement Measure Monogamous?\\}


\author{Yu Guo}
\email{guoyu3@aliyun.com}
\affiliation{Institute of Quantum Information Science, School of Mathematics and Statistics, Shanxi Datong University, Datong, Shanxi 037009, China}




\begin{abstract}
A crucial issue in quantum communication tasks is characterizing how quantum resources can be quantified and distributed over many parties.
Consequently, entanglement has been explored extensively.
However, the genuine entanglement still lacks of studying. 
There are few genuine multipartite entanglement measures and whether it is monogamous is unknown so far.
In this work, we explore the complete monogamy of genuine multipartite entanglement measure (GMEM) for which, at first,
we investigate a framework for unified/complete GMEM according to the unified/complete multipartite entanglement measure proposed in [Phys. Rev. A 101, 032301 (2020)].
We find a way of inducing unified/complete GMEM from any given unified/complete multipartite entanglement measure.
It is shown that any unified GMEM is completely monogamous, and any complete GMEM that induced by some given complete multipartite entanglement measure is tightly complete monogamous whenever the given complete multipartite entanglement measure is tightly complete monogamous.
In addition, the previous GMEMs are checked under this framework. It turns out that
the genuinely multipartite concurrence is not a good candidate as a GMEM.

\end{abstract}


\maketitle


\section{Introduction}

Entanglement is a quintessential manifestation
of quantum mechanics and is often considered to be a
useful resource for tasks like quantum teleportation or quantum
cryptography~\cite{Nielsen,Bennett1996prl,Horodecki2009,Guhne2009}, etc. 
There has been a
tremendous amount of research in the literatures aimed at
characterizing entanglement in the last three decades~\cite{Nielsen,Bennett1996prl,Horodecki2009,Guhne2009,Burkhart2021prx,Yuxiaodong2021pral,Luomx,Schmid,Navascues}.
In an effort to contribute to this
line of research, however, the genuine multiparty entanglement, 
which represents the strongest form of entanglement in many body
systems, still remains
unexplored or less studied in many facets.

A fundamental issue in this field is to quantify the genuine multipartite entanglement
and then analyze the distribution among the different parties. 
In 2000~\cite{Coffman}, Coffman \textit{et al}. presented a measure of genuine three-qubit
entanglement, called ``residual tangle'', and discussed the distribution relation for the first time.
In 2011, Ma \textit{et al}.~\cite{Ma2011} established postulates for a quantity to be
a GMEM and gave a genuine measure, called genuinely multipartite concurrence (GMC), by the origin bipartite concurrence.
The GMC is further explored in Ref.~\cite{Rafsanjani},
the generalized geometric measure is introduced in Refs.~\cite{Sen2010pra,Sadhukhan2017pra},
the average of ``residual tangle'' and GMC, i.e., $(\tau+C_{gme})/2$~\cite{Emary2004pra}, is shown to be genuine multipartite entanglement measures.
Another one is the divergence-based genuine multipartite entanglement measure presented in~\cite{Contreras-Tejada,Das}.
Recently, Ref.~\cite{Xie2021prl} introduced a new genuine three-qubit entanglement measure, called \textit{concurrence triangle}, which is quantified as the square root of the area of triangle deduced by concurrence. Consequently, 
we improved and supplemented the method in~\cite{Xie2021prl} and proposed a general way of defining GMEM in Ref.~\cite{G2021-2}.

The distribution of entanglement is believed to be monogamous, i.e., a quantum system entangled with
another system limits its entanglement with the remaining
others~\cite{Bennett1996pra}.
There are two ways in this research. The first one is analyzing monogamy relation based on bipartite entanglement measure,
and the second one is based on multipartite entanglement measure.
For the former one,
considerable efforts have been made in the last two
decades~\cite{Coffman,Osborne,streltsov2012are,Lan16,
	Ouyongcheng2007pra2,Deng,Karczewski,
	Bai,Koashi,
	Chengshuming,Allen,Hehuan,GG2019,GG,G2019,Camalet,
	Regula2016pra,Eltschka2015prl,Eltschka2018pra,Eltschka2019quantum,G2020}.
It is shown that almost all bipartite entanglement measures we known by now are monogamous.
In 2020, we established a framework for multipartite entanglement measure and discussed its monogamy relation which is called 
complete monogamy relation and tight complete monogamy relation~\cite{G2020}.
Under this framework, the distribution of entanglement becomes more clear since it displays a complete hierarchy relation
of different subsystems. We also proposed several multipartite entanglement measure and showed that they are completely monogamous.

The situation becomes much more complex when
we deal with genuine entanglement since it associates with not only multiparty system
but also the most complex entanglement structure. 
The main purpose of this work is to establish the framework of unified/complete GMEM, by which we then 
present the definition of complete monogamy and tight complete monogamy of unified and complete GMEM respectively.
Another aim is to find an approach of deriving GMEM from the multipartite entanglement measure introduced in
Ref.~\cite{G2020}.
In the next section we list some necessary concepts and the associated notations.
In Section III we discuss the framework of unified/complete GMEM and give several illustrated examples.
Then in Section IV, we investigate the complete monogamy relation and tight complete monogamy relation for GMEM accordingly.
A summary is concluded in the last section.

\section{Preliminary}

For convenience, in this section, we recall the concepts of genuine entanglement, 
complete multipartite entanglement measure, monogamy relation, complete monogamy relation, and genuine multipartite entanglement measure.
In the first subsection, we introduce the coarser relation of multipartite partition by which 
the following concepts can be easily processed.
For simplicity, throughout this paper, we denote by $\mH^{A_1A_2\cdots A_m}:=\mathcal{H}^{A_1}\otimes
\mathcal{H}^{A_2}\otimes\cdots\otimes\mathcal{H}^{A_m}$
an $m$-partite Hilbert space with finite dimension and by $\mS^{X}$ we denote the set of density
operators acting on $\mH^{X}$.

\subsection{Coarser relation of multipartite partition}

Let $X_1|X_2| \cdots |X_{k}$ be a partition (or called $k$-partition) of $A_1A_2\cdots A_m$, i.e., $X_s=A_{s(1)}A_{s(2)}\cdots A_{s(f(s))}$, $s(i)<s(j)$ whenever $i<j$, and $s(p)<t(q)$ whenever $s<t$ for any possible $p$ and $q$, $1\leq s,t\leq k$.
For instance, partition $AB|C|DE$ is a $3$-partition of $ABCDE$.
Let $X_1|X_2| \cdots |X_{k}$ and $Y_1|Y_2| \cdots |Y_{l}$ be two partitions of $A_1A_2\cdots A_n$ or subsystem of $A_1A_2\cdots A_n$.
$Y_1|Y_2| \cdots |Y_{l}$ is said to be \textit{coarser} than $X_1|X_2| \cdots |X_{k}$, denoted by 
\bea
X_1|X_2| \cdots| X_{k}\succ Y_1|Y_2| \cdots |Y_{l}, 
\eea
if $Y_1|Y_2| \cdots |Y_{l}$ can be obtained from $X_1|X_2| \cdots| X_{k}$
by one or two of the following ways (the coarser relation was also introduced in Ref.~\cite{Guo2021}, but  the third case in Ref.~\cite{Guo2021} is not valid here): 
\begin{itemize}
	\item (C1) Discarding some subsystem(s) of $X_1|X_2| \cdots| X_{k}$;
	\item (C2) Combining some subsystems of $X_1|X_2| \cdots| X_{k}$.
\end{itemize}
For example, $A|B|C|D|E \succ A|B|C|DE\succ A|B|C|D\succ AB|C|D\succ AB|CD$, $A|B|C|DE\succ A|B|DE$.
Clearly, $X_1|X_2| \cdots| X_{k}\succ Y_1|Y_2| \cdots |Y_{l}$ and $ Y_1|Y_2| \cdots |Y_{l}\succ Z_1|Z_2| \cdots| Z_{s}$ imply $X_1|X_2| \cdots| X_{k}\succ Z_1|Z_2| \cdots| Z_{s}$.

Furthermore, if $X_1|X_2| \cdots| X_{k}\succ Y_1|Y_2| \cdots |Y_{l}$,
we denote by $\Xi(X_1|X_2| \cdots| X_{k}- Y_1|Y_2| \cdots |Y_{l})$ the set of
all the partitions that are coarser than $X_1|X_2| \cdots| X_{k}$ and either exclude any subsystem of $Y_1|Y_2| \cdots |Y_{l}$ or include some but not all subsystems of $Y_1|Y_2| \cdots |Y_{l}$.
We take the five-partite system $ABCDE$ for example,
$\Xi(A|B|CD|E-A|B)=\left\lbrace CD|E, A|CD|E,B|CD|E,A|CD, A|E,
B|E,A|C,A|D,B|C,B|D\right\rbrace$.

For more clarity, we fix the following notations. Let $X_1|X_2| \cdots| X_{k}$ and $Y_1|Y_2| \cdots |Y_{l}$
be partitions of $A_1A_2\cdots A_n$ or subsystem of $A_1A_2\cdots A_n$.
We denote by 
\bea
X_1|X_2| \cdots| X_{k}\succ^a Y_1|Y_2| \cdots |Y_{l}
\eea 
for the case of (C1) 
and by,
\bea
X_1|X_2| \cdots| X_{k}\succ^b Y_1|Y_2| \cdots |Y_{l}
\eea 
for the case of of (C2).
For example, $A|B|C|D\succ^a A|B|D\succ^a B|D$,
$A|B|C|D\succ^b AC|B|D\succ^b AC|BD$.

\subsection{Multipartite entanglement}

An $m$-partite pure state $|\psi\rangle\in \mathcal{H}^{A_1A_2\cdots A_m}$ is called biseparable if it can be written as
$|\psi\rangle=|\psi\rangle^X \otimes |\psi\rangle^Y$
for some bipartition of $A_1A_2\cdots A_m$. 
$|\psi\ra$ is said to be $k$-separable if
$|\psi\ra=|\psi\ra^{X_1}|\psi\ra^{X_2}\cdots|\psi\ra^{X_k}$
for some $k$-partition of $A_1A_2\cdots A_m$.
$|\psi\ra$ is called fully separable if it is $m$-separable.
It is clear that whenever a
state is $k$-separable, it is automatically also $l$-separable for all $1<l<k\leq m$.
An $m$-partite mixed state $\rho$ is
biseparable if it can be written as a convex combination of
biseparable pure states
$\rho=\sum_{i}p_i|\psi_i\rangle \langle\psi_i|$, 
wherein the contained $\{|\psi_i\rangle\}$ can be biseparable with respect to different
bipartitions (i.e., a mixed biseparable state does not need to be separable with respect to any particular bipartition). Otherwise it is
called genuinely $m$-partite entangled (or called genuinely entangled briefly). 
We dnote by $\mS_g^{A_1A_2\cdots A_m}$ the set of
all genuinely entangled states in $\mS^{A_1A_2\cdots A_m}$.
Throughout this paper, for any $\rho\in\mS^{A_1A_2\cdots A_m}$ and any given $k$-partition $X_1|X_2|\cdots|X_k$
of $A_1A_2\cdots A_m$,
we denote by $\rho^{X_1|X_2|\cdots|X_k}$ the state for which we consider it as a $k$-partite state
with respect to the partition $X_1|X_2|\cdots|X_k$.

\subsection{Complete multipartite entanglement measure}

A function 
$E^{(m)}: \mS^{A_1A_2\cdots A_m}\to\bb{R}_{+}$ 
is called an $m$-partite entanglement measure in 
literatures~\cite{Horodecki2009,Hong2012pra,Hiesmayr2008pra} if
it satisfies:
\begin{itemize}
	\item {\bf(E1)} $E^{(m)}(\rho)=0$ if $\rho$ is fully separable;
	\item {\bf(E2)}
	$E^{(m)}$ cannot increase
	under $m$-partite LOCC.
\end{itemize}
An $m$-partite entanglement measure $E^{(m)}$ is said to be 
an $m$-partite entanglement monotone if it is convex and 
does not increase on average under $m$-partite stochastic LOCC.
For simplicity,
throughout this paper, if $E$ is an entanglement measure (bipartite, or multipartite) for pure states,
we define
\bea\label{eofmin}
E_F(\rho):=\min\sum_ip_iE^{(m)}(|\psi_i\ra)
\eea
and call it the convex-roof extension of $E$, where the minimum is taken over all pure-state
decomposition $\{p_i,|\psi_i\ra\}$ of $\rho$ (Sometimes, we use $E^F$ to denote $E_F$ hereafter). 
When we take into consideration an $m$-partite entanglement measure, we need 
discuss whether it is defined uniformly for any 
$k$-partite system at first, $k<m$. Let $E^{(m)}$ be a multipartite entanglement measure (MEM). If $E^{(k)}$ is uniquely determined by $E^{(m)}$ for any $2\leq k<m$, then we call $E^{(m)}$ a \textit{uniform} MEM. 
For example, GMC, denoted by $C_{gme}$~\cite{Ma2011}, is uniquely defined for any $k$, thus it is a uniform GMEM. Recall that,
\beax
C_{gme}(|\psi\ra):=\min\limits_{\gamma_i \in \gamma} \sqrt{2\left[ 1-\tr(\rho^{A_{\gamma_i}})^{2}\right] }
\eeax
for pure state $|\psi\ra\in\mH^{A_1A_2\cdots A_m}$, where $\gamma=\{\gamma_i\}$ represents the set of all possible bipartitions of $A_1A_2\cdots A_m$, and via the convex-roof extension for mixed states~\cite{Ma2011}.
All the unified MEMs presented in Ref.~\cite{G2020} are uniform MEM.
That is, a uniform MEM is series of MEMs that have uniform expressions definitely.
A uniform MEM $E^{(m)}$ is called a \textit{unified}
multipartite entanglement measure if it also satisfies the following
condition~\cite{G2020}: 
\begin{itemize}
	\item {\bf(E3)}~\textit{the unification condition}, i.e., 
	$E^{(m)}$ is consistent with $E^{(k)}$ for any $2\leqslant k<m$.
\end{itemize}
The unification condition should be comprehended in the following sense~\cite{G2020}.
Let 
$|\psi\ra^{A_1A_2\cdots A_m}=|\psi\ra^{A_1A_2\cdots A_k}|\psi\ra^{A_{k+1}\cdots A_m}$, then
\beax
E^{(m)}(|\psi\ra^{A_1A_2\cdots A_m})
=E^{(k)}(|\psi\ra^{A_1A_2\cdots A_k})+E^{(m-k)}|\psi\ra^{A_{k+1}\cdots A_m}.
\eeax
And
\beax
E^{(m)}(\rho^{A_1A_2\cdots A_m})=E^{(m)}(\rho^{\pi(A_1A_2\cdots A_m)})
\eeax 
for any $\rho^{A_1A_2\cdots A_m}\in\mS^{A_1A_2\cdots A_m}$, where $\pi$ is a permutation of the subsystems. 
In addition, 
\beax
E^{(k)}(X_1|X_2| \cdots| X_{k})\geqslant E^{(l)}(Y_1|Y_2| \cdots |Y_{l})
\eeax
for any $\rho^{A_1A_2\cdots A_m}\in\mS^{A_1A_2\cdots A_m}$ whenever $X_1|X_2| \cdots| X_{k}\succ^a Y_1|Y_2| \cdots |Y_{l}$,
where the vertical bar indicates the split
across which the entanglement is measured.
A uniform MEM $E^{(m)}$ is called a \textit{complete}
multipartite entanglement measure if it satisfies both {\bf(E3)} above and the following~\cite{G2020}: 
\begin{itemize}
	\item {\bf(E4)}~$E^{(m)}(X_1|X_2| \cdots| X_{k})\geqslant
	E^{(k)}(Y_1|Y_2| \cdots |Y_{l})$
	holds for all $\rho\in\mS^{A_1A_2\cdots A_m}$ whenever $X_1|X_2| \cdots| X_{k}\succ^b Y_1|Y_2| \cdots |Y_{l}$.
\end{itemize}
We need remark here that, although the partial trace is in fact a special trace-preserving completely positive map,
we cannot derive $\rho^{Y_1|Y_2| \cdots |Y_{l}}$ from $\rho^{X_1|X_2| \cdots| X_{k}}$
by any $k$-partite LOCC for any given $X_1|X_2| \cdots| X_{k}\succ Y_1|Y_2| \cdots |Y_{l}$.
Namely, different from that of bipartite case, the unification condition
can not induced by the $m$-partite LOCC.
For any bipartite measure $E$, $E(A|BC)\geq E(AB)$ for any $\rho^{ABC}$ since
$\rho^{AB}=\tr_C\rho^{ABC}$ can be obtained by partial trace on part $C$ and such a partial trace is in fact
a bipartite LOCC acting on $A|BC$. But $\rho^{AB}$ can not be derived from any tripartite LOCC acting on $\rho^{ABC}$. Thus, whether $E^{(3)}(A|BC)\geq E^{(2)}(AB)$ is unknown.

Several unified tripartite entanglement measures were proposed in Ref.~\cite{G2020}:
\beax
E_f^{(3)}\left( |\psi\rangle\right)&=&\frac12\left[S(\rho^A)+S(\rho^B)+S(\rho^C) \right],\\
\tau^{(3)}(|\psi\ra)&=&3- {\rm Tr}\left( \rho^A\right) ^2-{\rm Tr}\left( \rho^B\right) ^2-\tr\left( \rho^C\right) ^2,~~~\\
C^{(3)}(|\psi\ra)&=&\sqrt{\tau^{(3)}(|\psi\ra)},\\
N^{(3)}(|\psi\ra)&=&\tr^2 \sqrt{\rho^A}+\tr^2 \sqrt{\rho^B} +\tr^2 \sqrt{\rho^C} -3,\\
T^{(3)}_q(|\psi\ra)&=&\frac12\left[T_q(\rho^A)+T_q(\rho^B)+T_q(\rho^C) \right],~q>1,\\
R^{(3)}_{\alpha}(|\psi\ra)&=&\frac12R_{\alpha}(\rho^A\otimes\rho^B\otimes\rho^C),~0<\alpha<1
\eeax
for pure state $|\psi\ra\in\mH^{ABC}$,
and then by the convex-roof extension
for mixed state $\rho^{ABC}\in\mS^{ABC}$ (for mixed state, $N^{(3)}$ is replaced with $N_{F}^{(3)}$),
where
$T_q(\rho):=(1-q)^{-1}[\tr(\rho^q)-1]$
is the Tsallis $q$-entropy,
$R_{\alpha}(\rho):=(1-\alpha)^{-1}\ln(\tr\rho^\alpha)$
is the R\'{e}nyi $\alpha$-entropy.
In addition~\cite{G2020},
\be
N^{(3)}(\rho)=\|\rho^{T_a}\|_{\tr}+\|\rho^{T_b}\|_{\tr}+\|\rho^{T_c}\|_{\tr}-3
\ee
for any $\rho\in\mS^{ABC}$.
${E}_{f}^{(3)}$, $C^{(3)}$, $\tau^{(3)}$ and $T^{(3)}_{q}$ are shown to be complete tripartite entanglement measures
while $R^{(3)}_{\alpha}$, $N^{(3)}$ and $N_F^{(3)}$ are proved to be unified but not complete tripartite entanglement measures~\cite{G2020}.

In Ref.~\cite{Guo2020qip}, we introduce three unified tripartite entanglement measures (but not complete tripartite entanglement measures) in terms of fidelity:
\bea
E_{\mF}^{(3)}\left( |\psi\rangle\right)
&:=&1-\mF\left( |\psi\rangle\la\psi|,\rho^A\otimes\rho^B\otimes\rho^C\right),
~~\label{multipartite1}\\
E_{\mF'}^{(3)}\left( |\psi\rangle\right)
&:=&1-\sqrt{\mF}\left( |\psi\rangle\la\psi|,\rho^A\otimes\rho^B\otimes\rho^C\right),
~~\label{multipartite2}\\
E_{A\mF}^{(3)}\left( |\psi\rangle\right)
&:=&1-\mF_A\left( |\psi\rangle\la\psi|,\rho^A\otimes\rho^B\otimes\rho^C\right),
~~\label{multipartite3}
\eea
for any pure state $|\psi\ra$ in $\mH^{ABC}$,
where $\mF$ is the 
Uhlmann-Jozsa fidelity $\mF$~\cite{Jozsa1994,Uhlmann1976}, which is defined as
\bea
\mF(\rho,\sigma):=\left(  \tr\sqrt{\sqrt{\rho}\sigma\sqrt{\rho}}\right) ^2,
\eea
$\sqrt{\mF}$ is defined by~\cite{Zhanglin2015qip,Fawzi2015cmp,Luo2004pra}
\bea
\sqrt{\mF}(\rho,\sigma):=\sqrt{\mF(\rho,\sigma)},
\eea
and  the \emph{A-fidelity}, $\mF_A$, is the square of the quantum affinity
$A(\rho,\sigma)$~\cite{Ma2008pra,Raggio1984}, i.e.,
\bea
\mF_A(\rho,\sigma):=[\tr(\sqrt{\rho}\sqrt{\sigma})]^2.
\eea
For mixed states,
$E_{\mF,F}^{(3)}$,
$E_{\mF',F}^{(3)}$,
and $E_{A\mF,F}^{(3)}$ are defined by the convex-roof extension as in Eq.~\eqref{eofmin}.

\subsection{Monogamy relation}

For a given bipartite measure $Q$ (such as entanglement measure and other quantum correlation measure), $Q$ is said to be monogamous (we take the tripartite case for example) if~\cite{Coffman,Koashi}
\bea\label{monogamy1}
Q(A|BC)\geqslant Q(AB)+Q(AC).
\eea
However, Eq.~(\ref{monogamy1}) is not valid for many entanglement measures~\cite{Coffman,Zhuxuena2014pra,
	Bai,Luo2016pra} but some power function
of $Q$ admits the monogamy relation [i.e., $Q^\alpha(A|BC)\geqslant Q^\alpha(AB)+Q^\alpha(AC)$ for some $\alpha>0$].
In Ref.~\cite{GG}, we address this issue by proposing an improved definition of monogamy (without inequalities) for entanglement measure:
A bipartite measure of entanglement $E$ is monogamous if for any $\rho\in\mS^{ABC}$
that satisfies the \textit{disentangling condition}, i.e.,
\be\label{cond}
E( \rho^{A|BC}) =E( \rho^{AB}),
\ee
we have that $E(\rho^{AC})=0$, where $\rho^{AB}=\tr_C\rho^{ABC}$. With respect to this definition, a continuous measure $E$ is monogamous according to this definition if and only if
there exists $0<\alpha<\infty$ such that
\be\label{power}
E^\alpha( \rho^{A|BC}) \geqslant E^\alpha( \rho^{AB}) +E^\alpha( \rho^{AC})	
\ee
for all $\rho$ acting on the state space
$\mH^{ABC}$
with fixed $\dim\mH^{ABC}=d<\infty$ (see Theorem 1 in Ref.~\cite{GG}).
Notice that, for
these bipartite measures, only the
relation between $A|BC$, $AB$ and $AC$ are revealed, the global correlation
in ABC and the correlation contained in part $BC$ are missed~\cite{G2020}. That is, the monogamy relation in such a sense is not ``complete''. 
For a unified tripartite entanglement measure $E^{(3)}$,
it is said to be {\textit{completely monogamous}} if for any
$\rho\in\mathcal{S}^{ABC}$ that satisfies~\cite{G2020}
\be\label{condofm2}
E^{(3)}(\rho^{ABC})=E^{(2)}(\rho^{AB})
\ee
we have that $E^{(2)}(\rho^{AC})=E^{(2)}(\rho^{BC})=0$.
If $E^{(3)}$ is a continuous unified tripartite
entanglement measure. Then, $E^{(3)}$ is completely monogamous
if and only if there exists
$0<\alpha<\infty$ such that~\cite{G2020}
\bea\label{power2}
E^{\alpha}(\rho^{ABC})\geqslant  E^{\alpha}(\rho^{AB})
+ E^{\alpha}(\rho^{AC})+ E^{\alpha}(\rho^{BC})
\eea
for all $\rho^{ABC}\in\mathcal{S}^{ABC}$ with fixed $\dim\mH^{ABC}=d<\infty$,
here we omitted the superscript $^{(2,3)}$ of $E^{(2,3)}$ for brevity.
Let $E^{(3)}$ be a complete MEM. $E^{(3)}$ is defined to be
tightly complete monogamous if
for any state $\rho^{ABC}\in\mS^{ABC}$ that satisfying~\cite{G2020}
\bea\label{condofm5}
E^{(3)}(\rho^{ABC})=E^{(2)}(\rho^{A|BC})
\eea
we have $E^{(2)}(\rho^{BC})=0$,
which is equivalent to
\beax
E^{\alpha}(\rho^{ABC})\geqslant E^{\alpha}(\rho^{A|BC})+E^{\alpha}(\rho^{BC})
\eeax
for some $\alpha>0$, here we omitted the superscript $^{(2,3)}$ of $E^{(2,3)}$ for brevity.
For the general case of $E^{(m)}$, one can similarly followed with the same spirit.

\subsection{Genuine entanglement measure}

A function $E^{(m)}_{g}: \mS^{A_1A_2\cdots A_m}\to\bb{R}_{+}$ is defined to be a measure of genuine multipartite entanglement if it admits the following conditions~\cite{Ma2011}:
\begin{itemize}
	\item {\textbf{(GE1)}}~$E^{(m)}_{g}(\rho)=0$ for any biseparable $\rho\in\mS^{A_1A_2\cdots A_m}$.
	\item {\textbf{(GE2)}}~ $E^{(m)}_{g}(\rho)>0$ for any genuinely entangled state $\rho\in \mathcal{S}^{A_1A_2\cdots A_m}$.(This item can be weakened as: $E^{(m)}_{g}(\rho)\geqslant 0$ for any genuinely entangled state $\rho\in \mathcal{S}^{A_1A_2\cdots A_m}$. That is, maybe there exists some state which is genuinely entangled such that $E^{(m)}_{g}(\rho)= 0$. In such a case, the measure is called not faithful. Otherwise, it is called faithful. For example, the ``residual tangle'' is not faithful since it is vanished for the $W$ state.)
	\item {\textbf{(GE3)}}~ $E^{(m)}_{g}(\sum_ip_i\rho_i)\leqslant \sum_ip_iE^{(m)}_g(\rho_i)$ for any $\{p_i,\rho_i\}$, $\rho_i\in\mathcal{S}^{A_1A_2\cdots A_m}$, $p_i>0$, $\sum_ip_i=1$.
	\item {\textbf{(GE4)}}~ $E^{(m)}_{g}(\rho)\geqslant E^{(m)}_g(\rho')$ for any $m$-partite LOCC $\varepsilon$, $\varepsilon(\rho)=\rho'$.	
\end{itemize}
Note that \textbf{(GE4)} implies $E^{(m)}_{g}$ is invariant under local unitary transformations.
$E^{(m)}_{g}$ is said to be 
a genuine multipartite entanglement monotone if it  
does not increase on average under $m$-partite stochastic LOCC.
For example, $C_{gme}$ is a GMEM.

\section{Complete genuine multipartite entanglement measure}

Analogous to that of unified/complete multipartite entanglement measure
established in Ref.~\cite{G2020}, we discuss
the unification condition and the hierarchy condition for 
genuine multipartite entanglement measure in this section. 
We start out with observation of examples. 
Let $|\psi\rangle$ be an $m$-partite pure state in $\mathcal{H}^{A_1A_2\cdots A_m}$.
Recall that, the multipartite entanglement of formation ${E}_f^{(m)}$ is defined as~\cite{G2020}
\beax
{E}_f^{(m)}(|\psi\ra)&:=&\frac12\sum\limits_{i=1}^m S(\rho_{A_i}),
\eeax
where $\rho_X:={\rm Tr}_{\bar{X}}(|\psi\ra\la\psi|)$. 
We define
\bea
{E}_{g-f}^{(m)}(|\psi\ra)&:=&\frac12\delta(|\psi\ra)\sum\limits_{i=1}^m S(\rho_{A_i}), 
\eea
where $\delta(\rho)=0$ if $\rho$ is biseparable up to some bi-partition and $\delta(\rho)=1$ if $\rho$ is not biseparable up to any bi-partition.
For mixed state, it is defined by the convex-roof extension.
Obviously, ${E}_{g-f}^{(m)}$ is a uniform GMEM
since $I(A_1:A_2:\cdots :A_n)\geqslant 0$ for any $n$~\cite{Kumar2017pra}, where
$I(A_1:A_2:\cdots :A_n)
:=\sum_{k=1}^nS(\rho_{{A_k}})-S(A_1A_2\cdots A_n)
=S(\rho^{A_1A_2\cdots A_n}\|\rho^{A_1}\otimes\rho^{A_2}\otimes\cdots\rho^{A_n})
\geqslant0$.
The following properties are straightforward:
For any $\rho^{A_1A_2\cdots A_m}\in\mS_g^{A_1A_2\cdots A_m}$, 
\beax
{E}_{g-f}^{(k)}(X_1|X_2| \cdots| X_{k})>{E}_{g-f}^{(l)}(Y_1|Y_2| \cdots |Y_{l})
\eeax 
for any
	$X_1|X_2| \cdots| X_{k}\succ^b  Y_1|Y_2| \cdots |Y_{l}$.
It is worth noting that, for any uniform GMEM ${E}_g^{(m)}$, we cannot require ${E}_g^{(k)}(X_1|X_2|\cdots|X_{k})= {E}_g^{(l)}(Y_1|Y_2|\cdots|Y_{l})$
for any $\rho\in\mS_g^{A_1A_2\cdots A_m}$ and any $X_1|X_2| \cdots| X_{k}\succ^a Y_1|Y_2| \cdots |Y_{l}$.
For example, if ${E}_g^{(4)}(\rho^{ABCD})={E}_g^{(3)}(\rho^{ABC})$
for some $\rho^{ABCD}\in\mS_g^{ABCD}$, then the entanglement between part $ABC$ and part $D$ is zero, which means that $\rho^{ABCD}$ is biseparable with respect to the partition $ABC|D$, a contradiction. 
In addition, let $|\psi\ra^{ABC}$ be a tripartite genuine entangled state in $\mH^{ABC}$,
then $|\psi\ra^{ABC}|\psi\ra^D$ is not a four-partite genuine entangled state, i.e.,
\beax
E_g^{(4)}(|\psi\ra^{ABC}|\psi\ra^D)=0,
\eeax
but ${E}_g^{(3)}(\psi\ra^{ABC})>0$ provided that 
${E}_g^{(3)}$ is faithful.
That is, the genuine multipartite entanglement measure
is not necessarily decreasing under discarding of subsystem.
However, for the genuine entangled state, it is decreasing definitely.
From this observations, we give the following definition.

\begin{definition}
	Let $E_g^{(m)}$ be a uniform
	genuine entanglement measure. If it satisfies the \textit{unification condition}, i.e., 
	\bea 
	E_g^{(m)}({A_1A_2\cdots A_m})=E_g^{(m)}({\pi(A_1A_2\cdots A_m)})
	\eea
and 
		\bea\label{unification1}
		E_g^{(k)}(X_1|X_2|\cdots|X_{k})> E_g^{(l)}(Y_1|Y_2|\cdots|Y_{l})
		\eea
		for any $\rho\in\mS_g^{A_1A_2\cdots A_m}$ whenever $X_1|X_2| \cdots| X_{k}\succ^a Y_1|Y_2| \cdots |Y_{l}$,
	we call $E^{(m)}_{g}$ a \textit{unified} genuine multipartite entanglement measure, where $\pi(\cdot)$ denotes the permutation of the subsystems. 
\end{definition}

For any $\rho\in\mS_g^{A_1A_2\cdots A_m}$, if $X_1|X_2| \cdots| X_{k}\succ^b Y_1|Y_2| \cdots |Y_{l}$,  
We expect any unified GMEM 
satisfies ${E}_{g}^{(k)}(X_1|X_2|\cdots|X_{k})\geqslant {E}_{g}^{(l)}(Y_1|Y_2| \cdots |Y_{l})$ since `some amount of entanglement' may be hided in the combined subsystem.
For example, the quantity $E^{(3)}_g(AB|C|D)$ seems can not report the entanglement contained between subsystems $A$ and $B$.
We thus present the following definition.

\begin{definition}
Let $E_g^{(m)}$ be a unified GMEM.
If $E^{(m)}_{g}$ admits the \textit{hierarchy condition}, i.e.,
\bea\label{hierachy}
E_g^{(k)}(X_1|X_2|\cdots|X_{k})\geqslant E_g^{(l)}(Y_1|Y_2| \cdots |Y_{l})
\eea for any $\rho\in\mS_g^{A_1A_2\cdots A_m}$ whenever $X_1|X_2| \cdots| X_{k}\succ^b Y_1|Y_2| \cdots |Y_{l}$,
then it is said to be a \textit{complete} genuine multipartite entanglement measure.	
\end{definition}

By definition, ${E}_{g-f}^{(m)}$ is a complete GMEM. 
But $C_{gme}$ is not a complete GMEM since it does not satisfy the hierarchy condition~\eqref{hierachy}.	
We take a four-partite state for example.
Let 
\beax
|\psi\ra=\frac{\sqrt{5}}{4}|0000\ra+\frac14|1111\ra+\frac{\sqrt{5}}{4}|0100\ra+\frac{\sqrt{5}}{4}|1010\ra,
\eeax
then
$C_{gme}(|\psi\ra)=C(|\psi\ra^{ABC|D})=\frac{\sqrt{15}}{8}<C(|\psi\ra^{AB|CD})=\frac{\sqrt{65}}{8}$. 
In general, $C_{gme}$ is even not a unified GMEM since we can not guarantee the unification condition~\eqref{unification1} hold true.

We now turn to find unified/complete GMEM. ${E}_{g-f}^{(m)}$ is derived from unified/complete multipartite entanglement measures ${E}_f^{(m)}$. This motivates us to obtain unified/complete GMEMs
from the unified/complete MEMs.

\begin{pro}\label{gmem}
	Let $E^{(m)}$ be a unified/complete multipartite entanglement measure (resp. monotone), and define
	\bea\label{gmem1}
	E_{g-F}^{(m)}(\rho):=\min_{\{p_i,|\psi_i\ra\}}\sum p_i\delta(|\psi_i\ra)E^{(m)}(|\psi_i\ra)
	\eea
	whenever $E_F^{(m)}=\min_{\{p_i,|\psi_i\ra\}}\sum p_iE^{(m)}(|\psi_i\ra)$
	and 
	\bea\label{gmem2}
	E_g^{(m)}(\rho):=\delta(\rho)E^{(m)}(\rho)
	\eea
	whenever $E^{(m)}$ is not defined by the convex-roof extension for mixed state,
	where the minimum is 
	taken over all pure-state
	decomposition $\{p_i,|\psi_i\ra\}$ of $\rho\in\mS^{A_1A_2\cdots A_m}$,
	$\delta(\rho)=1$ whenever $\rho$ is genuinely entangled and $\delta(\rho)=0$ otherwise.
	Then $E_g^{(m)}$ is a unified/complete genuine multipartite entanglement measure (resp. monotone). 
\end{pro}

\begin{proof}
It is clear that $E_{g-F}^{(m)}$ and $E_g^{(m)}$ satisfy the unification condition (resp. hierarchy condition) on 
$\mS_g^{A_1A_2\cdots A_m}$ whenever $E^{(m)}$ satisfies the unification condition (resp. hierarchy condition) on 
$\mS^{A_1A_2\cdots A_m}$.
\end{proof}

Consequently, according to Proposition~\ref{gmem},
we get 
\beax
\tau_g^{(3)}(|\psi\ra)&=&\delta(|\psi\ra)\left[  3- {\rm Tr}\left( \rho^A\right) ^2-{\rm Tr}\left( \rho^B\right) ^2-\tr\left( \rho^C\right) ^2\right] ,~~~\\
C_g^{(3)}(|\psi\ra)&=&\sqrt{\tau_g^{(3)}(|\psi\ra)},\\
N_g^{(3)}(|\psi\ra)&=&\delta(|\psi\ra)\left[\tr^2 \sqrt{\rho^A}+\tr^2 \sqrt{\rho^B} +\tr^2 \sqrt{\rho^C} -3\right], \\
T^{(3)}_{g-q}(|\psi\ra)&=&\frac12\delta(|\psi\ra)\left[T_q(\rho^A)+T_q(\rho^B)+T_q(\rho^C) \right],~q>1,\\
R^{(3)}_{g-\alpha}(|\psi\ra)&=&\frac12\delta(|\psi\ra)R_{\alpha}(\rho^A\otimes\rho^B\otimes\rho^C),~0<\alpha<1,\\
E_{g-\mF}^{(3)}\left( |\psi\rangle\right)
&=&\delta(|\psi\ra)\left[ 1-\mF\left( |\psi\rangle\la\psi|,\rho^A\otimes\rho^B\otimes\rho^C\right)\right] ,\\
E_{g-\mF'}^{(3)}\left( |\psi\rangle\right)
&=&\delta(|\psi\ra)\left[1-\sqrt{\mF}\left( |\psi\rangle\la\psi|,\rho^A\otimes\rho^B\otimes\rho^C\right)\right] ,\\
E_{g-A\mF}^{(3)}\left( |\psi\rangle\right)
&=&\delta(|\psi\ra)\left[1-\mF_A\left( |\psi\rangle\la\psi|,\rho^A\otimes\rho^B\otimes\rho^C\right)\right],
\eeax
for pure states and define by the convex-roof extension for the mixed states (for mixed state, $N_g^{(3)}$ is replaced with the convex-roof extension of $N_g^{(3)}$, $N_{g-F}^{(3)}$),
and  
\beax
N_g^{(3)}(\rho)=\delta(\rho)\left( \|\rho^{T_a}\|_{\tr}+\|\rho^{T_b}\|_{\tr}+\|\rho^{T_c}\|_{\tr}-3\right) 
\eeax
for any $\rho\in\mS^{ABC}$.
These tripartite measures, except for $N_g^{(3)}$ are in fact special cases of $\mE^F_{g-123}$ in Ref.~\cite{G2021-2}.
Generally, we can define 
\beax
\tau_g^{(m)}(|\psi\ra)&=&\delta(|\psi\ra)\left[  m- \sum\limits_{i}{\rm Tr}\left( \rho^{A_i}\right) ^2\right] ,~~~\\
C_g^{(m)}(|\psi\ra)&=&\sqrt{\tau_g^{(m)}(|\psi\ra)},\\
N_g^{(m)}(|\psi\ra)&=&\delta(|\psi\ra)\left[\sum\limits_{i}\tr^2 \sqrt{\rho^{A_i}}-m\right], \\
T^{(m)}_{g-q}(|\psi\ra)&=&\frac12\delta(|\psi\ra)\sum\limits_{i}T_q(\rho^{A_i}),~q>1,\\
R^{(m)}_{g-\alpha}(|\psi\ra)&=&\frac12\delta(|\psi\ra)R_{\alpha}\left( \bigotimes\limits_{i}\rho^{A_i}\right) ,~0<\alpha<1,
\eeax
\beax
E_{g-\mF}^{(m)}\left( |\psi\rangle\right)
&=&\delta(|\psi\ra)\left[ 1-\mF\left( |\psi\rangle\la\psi|,\bigotimes\limits_{i}\rho^{A_i}\right)\right] ,\\
E_{g-\mF'}^{(m)}\left( |\psi\rangle\right)
&=&\delta(|\psi\ra)\left[1-\sqrt{\mF}\left( |\psi\rangle\la\psi|,\bigotimes\limits_{i}\rho^{A_i}\right)\right] ,\\
E_{g-A\mF}^{(m)}\left( |\psi\rangle\right)
&=&\delta(|\psi\ra)\left[1-\mF_A\left( |\psi\rangle\la\psi|,\bigotimes\limits_{i}\rho^{A_i}\right)\right],
\eeax
for pure states and define by the convex-roof extension for the mixed states (for mixed state, $N_g^{(m)}$ is replaced with $N_{g-F}^{(m)}$),
and  
\beax
N_g^{(m)}(\rho)=\delta(\rho)\left( \left\| \sum\limits_{i}\rho^{T_i}\right\| _{\tr}-m\right) 
\eeax
for any $\rho\in\mS^{A_1A_2\cdots A_m}$.
According to Proposition~\ref{gmem}, together with Theorem 5 in Ref.~\cite{G2020}, the statement below is straightforward.

\begin{pro}
	${E}_{g-f}^{(m)}$, $\tau_g^{(m)}$,
	$C_g^{(m)}$, and $T_{g-q}^{(m)}$ are complete genuine multipartite entanglement monotones while
	$R_{g-\alpha}^{(m)}$, $N_{g-F}^{(m)}$, $N_g^{(m)}$, $E_{g-\mF}^{(m)}$, $E_{g-\mF'}^{(m)}$, and $E_{g-A\mF}^{(m)}$ are unified genuine multipartite entanglement monotones but not complete genuine multipartite entanglement monotones. 
\end{pro}

Very recently, we proposed the following genuine four-partite entanglement measures~\cite{G2021-2}.
Let $E$ be a bipartite entanglement measure and let
\bea\label{1234(2)-2}
\mE_{g-1234(2)}(|\psi\ra):=\delta(|\psi\ra) \sum_ix_i^{(2)}
\eea
for any given $|\psi\ra\in\mH^{ABCD}$,
where 
$E(|\psi\ra^{AB|CD})=x_1^{(2)}$,
$E(|\psi\ra^{A|BCD})=x_2^{(2)}$,
$E(|\psi\ra^{AC|BD})=x_3^{(2)}$,
$E(|\psi\ra^{ABC|D})=x_4^{(2)}$,
$E(|\psi\ra^{AD|BC})=x_5^{(2)}$,
$E(|\psi\ra^{B|ACD})=x_6^{(2)}$,
$E(|\psi\ra^{C|ABD})=x_7^{(2)}$.
Then $\mE^F_{g-1234(2)}$ is a genuine four-partite entanglement measure.
Let $E^{(3)}$ be a tripartite entanglement measure, 
\bea
\mE_{g-1234(3)}(|\psi\ra)=\delta(|\psi\ra)\sum_ix_i^{(3)}
\eea
for any given $|\psi\ra\in\mS^{ABCD}$,
where
$E^{(3)}(\rho^{A|B|CD})=x_1^{(3)}$,
$E^{(3)}(\rho^{A|BC|D})=x_2^{(3)}$,
$E^{(3)}(\rho^{AC|B|D})=x_3^{(3)}$,
$E^{(3)}(\rho^{AB|C|D})=x_4^{(3)}$,
$E^{(3)}(\rho^{AD|B|C})=x_5^{(3)}$,
$E^{(3)}(\rho^{A|BD|C})=x_6^{(3)}$.
It is clear that $\mE^F_{g-1234(3)}$ is a genuine four-partite entanglement measures but not uniform GMEM.

Generally, we can define
$\mE^F_{g-1234\cdots m(2)}$ by the same way and it is a uniform GMEM.
We check below $\mE^F_{g-1234\cdots m(2)}$ is a complete GMEM whenever $E$ is an entanglement monotone.
We only need to discuss the case of $m=4$ and the general cases can be argued similarly.
For any genuine entangled pure state $|\psi\ra\in\mH^{ABCD}$, and any bipartite entanglement monotone $E$,
it is clear that $\mE_{g-1234(2)}(|\psi\ra)>E^F(\rho^{XY})$ for any 
$\{X,Y\}\in\{A,B,C,D\}$.
For any pure state decomposition of $\rho^{ABC}$,
$\rho^{ABC}=\sum_ip_i|\psi_i\ra\la\psi_i|$,
we have $E(|\psi\ra^{A|BCD})\geqslant\sum_ip_iE(|\psi_i\ra^{A|BC})$,
$E(|\psi\ra^{AB|CD})\geqslant\sum_ip_iE(|\psi_i\ra^{AB|C})$,
and $E(|\psi\ra^{B|ACD})\geqslant\sum_ip_iE(|\psi_i\ra^{B|AC})$
since any ensemble $\{p_i,|\psi_i\ra\}$ can be derived by LOCC from $|\psi\ra$.
It follows that
$\mE_{g-1234(2)}(|\psi\ra)>\mE^F_{g-123(2)}(\rho^{ABC})$.
By symmetry of the subsystems, we get the unification condition is valid for pure state.
For mixed state $\rho\in\mS_g^{ABCD}$,
we let 
\beax
\mE^F_{g-1234(2)}(\rho)=\sum_jp_j\mE_{g-1234(2)}(|\phi_j\ra)
\eeax
for some decomposition $\rho=\sum_jp_j|\phi_j\ra\la\phi_j|$.
Then
\beax
\mE_{g-1234(2)}(|\phi_j\ra)\geqslant\mE^F_{g-123(2)}(\rho_j^{ABC})
\eeax
for any $j$, where $\rho_j^{ABC}=\tr_D(|\phi_j\ra\la\phi_j|)$.
Therefore
\beax
\mE^F_{g-1234(2)}(\rho)=\sum_jp_j\mE_{g-1234(2)}(|\phi_j\ra)
\geqslant\sum_jp_j\mE^F_{g-123(2)}(\rho_j^{ABC})
\geqslant\mE^F_{g-123(2)}(\rho^{ABC})
\eeax
as desired.
In addition, it is clear that
\bea
\mE^F_{g-123(2)}(\rho^{ABC})>E^F(\rho^{AB})
\eea 
for any $\rho\in\mS_g^{ABCD}$. That is,
$\mE^F_{g-1234\cdots m(2)}$ is a unified GMEM.
The hierarchy condition is obvious.
Thus $\mE^F_{g-1234\cdots m(2)}$ is a complete GMEM whenever $E$ is an entanglement monotone.

\begin{remark}\label{remark1}
	It is clear that, for $\mE^F_{g-1234\cdots m(2)}$, the inequality in Eq.~\eqref{hierachy} is a strict inequality, i.e., 
	\bea\label{hierachy2}
	E_g^{(k)}(X_1|X_2|\cdots|X_{k})> E_g^{(l)}(Y_1|Y_2| \cdots |Y_{l})
	\eea for any $\rho\in\mS_g^{A_1A_2\cdots A_m}$ whenever $X_1|X_2| \cdots| X_{k}\succ^b Y_1|Y_2| \cdots |Y_{l}$.
	In addition, according to the proof of Proposition 4 in Ref.~\cite{G2020},
	Eq.~\eqref{hierachy} holds for ${E}_{g-f}^{(m)}$, $\tau_g^{(m)}$,
	$C_g^{(m)}$, and $T_{g-q}^{(m)}$.
	Namley, in general, there does not exist $\rho\in\mS_g^{A_1A_2\cdots A_m}$ such that $E_g^{(k)}(X_1|X_2|\cdots|X_{k})=E_g^{(l)}(Y_1|Y_2| \cdots |Y_{l})$ holds, $X_1|X_2| \cdots| X_{k}\succ^b Y_1|Y_2| \cdots |Y_{l}$.
\end{remark}

\section{Complete monogamy of genuine multipartite entanglement measure}

We are now ready for discussing the complete monogamy relation of GMEM.
By the previous arguments, 
the genuine multipartite entanglement measure
is not necessarily decreasing under discarding of subsystem.
However, for the genuine entangled state, it does decrease.
We thus conclude the following definition of complete monogamy for genuine entanglement measure.

\begin{definition}
	Let $E_g^{(m)}$ be a uniform GMEM. We call $E_g^{(m)}$
	is completely monogamous if for any $\rho\in\mS_g^{A_1A_2\cdots A_m}$ we have 
	\bea
	E_g^{(k)}\left(\rho^{X_1|X_2|\cdots|X_{k}}\right)>E_g^{(l)}\left(\rho^{Y_{1}|Y_{2}|\cdots|Y_{l}}\right) 
	\eea
	holds for all $X_1|X_2| \cdots| X_{k}\succ^a Y_{1}|Y_{2}|\cdots|Y_{l}$.	
\end{definition}

That is, any unified GMEM is completely monogamous.
Moreover, according to the proof of Theorem 1 in Ref.~\cite{GG}, we can get the equivalent statement of complete monogamy for continuous genuine tripartite entanglement measure (the general $m$-partite case can be followed in the same way).

\begin{pro}\label{monogamypower}
	Let $E_g^{(3)}$ be a continuous uniform genuine tripartite 
	entanglement measure. Then, $E_g^{(3)}$ is completely monogamous
	if and only if there exists
	$0<\alpha<\infty$ such that
	\bea\label{power3}
	E_g^{\alpha}(\rho^{ABC})>  E^{\alpha}(\rho^{AB})
	+ E^{\alpha}(\rho^{AC})+ E^{\alpha}(\rho^{BC})
	\eea
	for all $\rho^{ABC}\in\mathcal{S}_g^{ABC}$ with fixed $\dim\mH^{ABC}=d<\infty$,
	here we omitted the superscript $^{(3)}$ of $E^{(3)}$ for brevity.
\end{pro}

Analogously, for the four-partite case, if $E_g^{(4)}$ is a continuous uniform GMEM,
then $E_g^{(4)}$ is completely monogamous
if and only if there exist
$0<\alpha, \beta<\infty$ such that
\bea\label{power23}
E_g^{\alpha}(\rho^{ABCD})&>&  E_g^{\alpha}(\rho^{ABC})
+ E_g^{\alpha}(\rho^{ABD})+E_g^{\alpha}(\rho^{ACD})
+ E_g^{\alpha}(\rho^{BCD}),\\
E_g^{\beta}(\rho^{ABCD})&>&  E^{\beta}(\rho^{AB})
+ E^{\beta}(\rho^{BC})+E^{\beta}(\rho^{AC})+ E^{\beta}(\rho^{BD})+E^{\beta}(\rho^{AD})
+ E^{\beta}(\rho^{CD})~~~~
\eea
for all $\rho^{ABCD}\in\mathcal{S}_g^{ABCD}$ with fixed $\dim\mH^{ABC}=d<\infty$,
here we omitted the superscript $^{(3,4)}$ of $E^{(3,4)}$ for brevity.
Since $C_{gme}$ may be not a unified GMEM,
we conjecture that $C_{gme}$ is not completely monogamous.

As a counterpart to the tightly complete monogamous relation of the complete multipartite entanglement measure in Ref.~\cite{G2020},
we give the following definition.

\begin{definition}\label{tight-monogamy}
	Let $E_g^{(m)}$ be a complete GMEM. We call $E_g^{(m)}$
	is tightly complete monogamous if it satisfies the genuine disentangling condition, i.e.,
	either for any $\rho\in\mS_g^{A_1A_2\cdots A_m}$ that satisfies 
	\bea\label{hierachy-equation}
	E_g^{(k)}({X_1|X_2|\cdots|X_{k}})=E_g^{(l)}({Y_1|Y_2|\cdots|Y_{l}}) 
	\eea
	we have that
	\bea
	E_g^{(\ast)}({\Gamma}) =0
	\eea
	holds for all $\Gamma\in \Xi(X_1|X_2| \cdots| X_{k}- Y_1|Y_2| \cdots |Y_{l})$, 
	or 
	\bea\label{tight-2}
	E_g^{(k)}({X_1|X_2|\cdots|X_{k}})>E_g^{(l)}({Y_1|Y_2|\cdots|Y_{l}}) 
	\eea
	holds for any $\rho\in\mS_g^{A_1A_2\cdots A_m}$, where $X_1|X_2| \cdots| X_{k}\succ^b Y_1|Y_2| \cdots |Y_{l}$,
	and the superscript $(\ast)$ is associated with the partition $\Gamma$, e.g., if $\Gamma$ is a $n$-partite partition, then $(\ast)=(n)$.
\end{definition}

\begin{remark}
	According to Remark~\ref{remark1}, for ${E}_{g-f}^{(m)}$, $\tau_g^{(m)}$,
	$C_g^{(m)}$, $T_{g-q}^{(m)}$, and $\mE^F_{g-1234\cdots m(2)}$, the case of Eq.~\eqref{hierachy-equation}
	can not occur, so they are tightly complete monogamous.
	We conjecture that the case of Eq.~\eqref{hierachy-equation}
	can not occur for any complete GMEM.
	In such a sense, any complete GMEM is tightly complete monogamous. 
\end{remark}

For example, if $E_g^{(3)}$ is a complete GMEM, then
$E_g^{(3)}$
is tightly complete monogamous if 
for any $\rho^{ABC}\in\mS_g^{ABC}$ that satisfying
\bea\label{condofm4}
E_g^{(3)}(\rho^{ABC})=E^{(2)}(\rho^{A|BC})
\eea
we have $E^{(2)}(\rho^{BC})=0$, and $E_g^{(3)}$
is completely monogamous
\bea\label{condofm3}
E_g^{(3)}(\rho^{ABC})>E^{(2)}(\rho^{AB})
\eea
is always correct for any $\rho^{ABC}\in\mS_g^{ABC}$.	
That is, the complete monogamy of ${E}_{g}^{(m)}$ refers to it is completely monogamous on genuine entangled state,
and ${E}_{g}^{(m)}$ is strictly decreasing under discarding of subsystem,
which is different from that of complete entanglement measure.
Equivalently, if $E_g^{(3)}$ is a continuous complete GMEM, 
then
$E_g^{(3)}$
is tightly complete monogamous if and only if there exists
$0<\alpha<\infty$ such that
\bea\label{power5}
E_g^{\alpha}(\rho^{ABC})\geqslant  E^{\alpha}(\rho^{AB})
+ E^{\alpha}(\rho^{AB|C})
\eea
holds for all $\rho^{ABC}\in\mathcal{S}_g^{ABC}$ with fixed $\dim\mH^{ABC}=d<\infty$,
here we omitted the superscript $^{(3)}$ of $E^{(3)}$ for brevity.

By definition~\ref{tight-monogamy}, $\mE^F_{g-1234\cdots m(2)}$ is tightly complete monogamous
since for $\mE^F_{g-1234\cdots m(2)}$ the genuine disentangling condition~(\ref{tight-2})
always holds. $C_{gme}$ is not tightly complete monogamous since it violates the genuine disentangling condition.
In addition, the tightly complete monogamy of ${E}_{g}^{(m)}$ is closely related to that of ${E}^{(m)}$
whenever ${E}_{g}^{(m)}$ is derived from ${E}^{(m)}$ as in Eqs.~(\ref{gmem1}) or (\ref{gmem2}).

\begin{pro}
	Let ${E}^{(m)}$ be a complete multipartite entanglement measure. If ${E}^{(m)}$ is tightly complete monogamous, then the genuine multipartite entanglement measure ${E}_{g}^{(m)}$, induced by ${E}^{(m)}$ as in Eqs.~(\ref{gmem1}) or (\ref{gmem2}), is tightly complete monogamous.	
\end{pro}

Together with Proposition 4 in Ref.~\cite{G2020}, 
$R_{g-\alpha}^{(m)}$, $N_{g-F}^{(m)}$ and $N_g^{(m)}$ are completely monogamous but not tightly complete monogamous.

\section{Conclusion and discussion}

We have proposed a framework of unified/complete genuine multipartite entanglement measure, 
from which we established the scenario of complete monogamy and tightly complete monogamy of genuine multipartite entanglement measure.
The spirit here is consistent with that of unified/complete multipartite entanglement measure in Ref.~\cite{G2020}.
We also find a simple way of deriving unified/complete genuine multipartite entanglement measure from the unified/complete multipartite entanglement measure.
Under such a framework, the multipartite entanglement becomes more clear, and, in addition, we can judge whether a given genuine entanglement measure is good or not.
Comparing with other multipartite entanglement measure, the unified genuine entanglement measure
is completely monogamous automatically.
That is, genuine entanglement display the monogamy of entanglement more evidently than other measures.
These results support that entanglement is monogamous as we expect.
We thus suggest that, monogamy should be a necessary requirement for a genuine entanglement measure.

\begin{acknowledgements}
This work is supported by the National Natural Science Foundation of
China under Grant No.~11971277, the Fund Program for the Scientific Activities of Selected Returned Overseas Professionals in Shanxi Province, and the Scientific Innovation Foundation of the Higher 
Education Institutions of Shanxi Province under Grant No.~2019KJ034. 
\end{acknowledgements}




\end{document}